\documentstyle[aps,pre,twocolumn,floats,epsfig]{revtex}

\newcommand{\equ}[1]{(\protect\ref{#1})}

\begin{document}
\draft
\wideabs{
\title{Model of correlated sequential adsorption of colloidal
  particles} 
 
\author{R. Pastor-Satorras$^1$ and J. M. Rub{\'\i}$^2$}

\address{$^1$Dept. de F{\'\i}sica i Enginyeria Nuclear,
  Universitat Polit{\`e}cnica de Catalunya,\\
  Campus Nord, M{\`o}dul B4,  08034 Barcelona, Spain\\
  $^2$Dept. de F{\'\i}sica Fonamental, Facultat de
  F{\'\i}sica --- CER F{\'\i}sica Sistemes Complexos,\\
  Universitat de Barcelona, Av. Diagonal 647, 08028 Barcelona, Spain}
\date{\today} 

\maketitle

\begin{abstract}
  We present results of a new model of sequential adsorption in which
  the adsorbing particles are correlated with the particles attached
  to the substrate. The strength of the correlations is measured by a
  tunable parameter $\sigma$. The model interpolates between free ballistic
  adsorption in the limit $\sigma\to\infty$ and a strongly correlated phase,
  appearing for $\sigma\to0$ and characterized by the emergence of highly
  ordered structures. The phenomenon is manifested through the
  analysis of several magnitudes, as the jamming limit and the
  particle-particle correlation function. The effect of correlations
  in one dimension manifests in the increased tendency to particle
  chaining in the substrate. In two dimensions the correlations induce
  a percolation transition, in which a spanning cluster of connected
  particles appears at a certain critical value $\sigma_c$.  Our study
  could be applicable to more general situations in which the coupling
  between correlations and disorder is relevant, as for example, in
  the presence of strong interparticle interactions.
\end{abstract}

}

\section{Introduction}

The study of the irreversible adsorption of colloidal particles onto a
surface has attracted a great deal of interest in the last years, due
to its many practical applications in physics, chemistry, biophysics,
medicine, etc. \cite{bartelt91,evans93}. The understanding of these
processes has been deepened mainly through the formulation of
different models, defined via a set of rules by which the particles
accommodate upon arriving at the surface.  The different rules are
responsible for the different values of the quantities describing the
geometry of the absorbed phase, as the maximum fraction of the surface
covered by particles---the {\em jamming limit} $\theta_\infty$---or the
particle-particle pair correlation function $g(r)$.  In the random
sequential adsorption model (RSA)
\cite{renyi63,feder80,schaaf88,senger91,ramsden93} particles are
placed at randomly selected positions on the surface.  When an
incoming particles overlaps with a previously adsorbed one, it is
rejected; otherwise, it becomes irreversibly adsorbed. The RSA model
is thus a good approximation when particles arrive at the surface
purely by diffusion, and excluded volume effects are predominant
\cite{senger91}. In the ballistic model (BM)
\cite{meakin87,talbot92,jullien92,thompson92}, on the other hand,
particles descend to the surface following straight vertical
trajectories. An incoming particle that does not reach the surface
directly is allowed to roll over the previously adsorbed ones,
following the steepest descent path, until it reaches a stable
position. Particles that eventually rest on the surface are
irreversibly adsorbed; otherwise they are rejected. The BM is thus a
valid approximation to describe adsorption in the presence of very
strong interactions between particles and substrate
\cite{senger93,schaaf95,pastor98a}.

In the models described above, and in many of the variations of them
analyzed so far, particles are supposed to interact only through
short-range interactions---hard-core repulsion. The main exceptions
are the analysis of the role played by electrostatic \cite{adamczyk96}
and dipolar \cite{pastor98a} interactions. In the absence of
long-range interactions, the particles arriving at the surface are
essentially uncorrelated from the adsorbed phase, and interact with
them only via excluded-volume effects. Therefore, little is known
about the general effect of correlations among the adsorbed phase and
the adsorbing particles \cite{notecorrelations}.

Our purpose in this paper is to analyze the influence of correlations
among particles in the structure of the adsorbed phase in a simple
numerical model. In our model, particles adsorb sequentially onto the
surface. The presence of correlations affects the position on the
surface in which the particles try to adsorb.  In order to mimic the
effect of long-range interactions, which tend to attract the incoming
particles in the vicinity of the particles already attached, the trial
position of the next particle is selected to be the position of a
particle in the substrate plus an increment $\xi$, selected at random
from a given probability distribution $\rho(\xi)$.  Once the trial position
has been selected, the adsorption process proceeds according to the
rules of the BM model. The distribution $\rho(\xi)$ depends on a parameter
$\sigma$ that interpolates between a flat distribution and a delta
function, which allows to explore the range between absence of
correlations and a very strong correlation effect, respectively.
Apart from the obvious relation with the adsorption in presence of
long-range attractive interactions \cite{pastor98a}, the inclusion of
general correlations in adsorption phenomena is relevant in other kind
of related problems, such as car parking, bird nesting, or adsorption
with memory effects, where the state of the adsorbed phase at a given
time exerts a strong influence on the position of the next incoming
unit. We want to note the our model differs from the generalized
ballistic-deposition models presented in Refs.~\cite{viot93,choi95}
due to the implicit role as nucleation centers played in our model by
the adsorbed particles.

The paper is organized as follows. In Section II we introduce the
model in a general context, whereas in Sec. III we present the results
of numerical simulations in dimensions $d=1$ and $2$. Finally, in the
last Section we discuss the common features induced by correlations in
the structure and geometry of the adsorbed phase.

\begin{figure}[t]
  \centerline{\epsfig{file=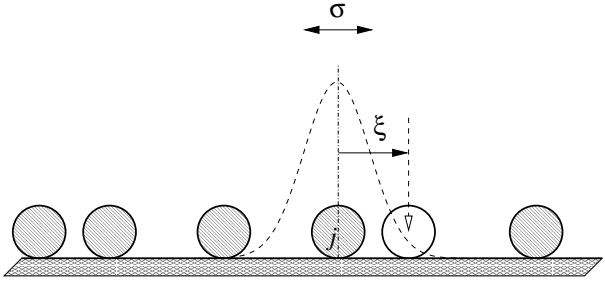,width=8cm}}
  \vspace*{0.2cm}
  \caption{Adsorption rule for the correlated adsorption model. An
    adsorbed particle $j$ is randomly selected. The newly adsorbed
    particle is deposited at a distance $\xi$ from the center of the
    selected particle. If it does not overlap with any other adsorbed
    particle, it is irreversibly attached. Otherwise, it is adsorbed
    following the prescription of the BM deposition.}
  \label{fig:landing}
\end{figure}

\section{Model}

Our model considers the sequential adsorption of particles of diameter
$a$ onto a $d$-dimensional surface; it is thus $d+1$ dimensional,
including the direction perpendicular to the plane in which the
particles move before adsorption. Particles arrive at the surface
following straight vertical trajectories, and upon touching the
surface or any adsorbed particle, they are attached following the
rules of the BM. The key ingredient in the model, that differs from
previous works, is the selection of the landing point of the arriving
particles. In order to introduce correlations among the incoming
particles and the adsorbed phase, the landing point is selected as
follows: The first particle is locate at random on the substrate.
Suppose that, at a given time step, the substrate is composed by set
of $N$ particles, located at positions ${\mathbf R}_i$, $i=1,\ldots N$. The
landing position of the next particle is chosen as
\begin{equation}
  \label{eq:landing}
  {\mathbf R}' = {\mathbf R}_j +  {\mathbf \hat{n}} \, \xi,
\end{equation}
where ${\mathbf R}_j$ is the position on the surface of a particle $j$
randomly selected among the $N$ particles present on the adsorbed
phase, ${\mathbf \hat{n}}$ is a unity vector, parallel to the
adsorption plane and oriented at random, and the distance $\xi$ is
Gaussian random variable, distributed according to the density
\begin{equation}
  \label{eq:gaussian}
  \rho(\xi) = \frac{1}{\sqrt{2 \pi} a \sigma} \exp\left( -
  \frac{\xi^2}{2 a^2 \sigma^2} \right),
\end{equation}
see Fig.~\ref{fig:landing}. The only parameter in the model is $\sigma$,
defined as the root-mean-square distance from an adsorbed particle at
which the new particle is probed. Once selected the position ${\mathbf
  R}'$, the particle is attached following the rules of the BM. If the
particle is accepted, the updated position ${\mathbf R}''$ becomes the
new value ${\mathbf R}_{N+1}$. If the particle is rejected, a new
position is chosen and the procedure iterated.

The above algorithm has a very intuitive interpretation. Each one of
the particles in the substrate play the role of a {\em nucleation
  center}, attracting incoming particles towards them. We can envision
this process as the result of long-range attractive interactions
between the elements of the adsorbed phase and the incoming particles.
The attraction has a random component given by the Gaussian term in
Eq.~\equ{eq:landing}, which represents the effect of the forces
exerted on the incoming particle by the regions of the substrate
located far away from the landing point.  The strength of the
attraction is given by the parameter $\sigma$ in Eq.~\equ{eq:gaussian}.
With this parameter we can interpolate between absolute lack of
correlations, achieved in the limit $\sigma\to\infty$, and the case of very strong
correlations, in the limit $\sigma\to 0$. In the former case, the Gaussian
\equ{eq:gaussian} tends to a flat distribution, which corresponds to a
free BM adsorption. On the other hand, in the limit $\sigma\to 0$ the
distribution \equ{eq:gaussian} becomes a sharply peaked delta
function. In this case the correlations are maximal, and one would
expect to obtain a close-packed substrate, at least in $1+1$
dimensions.

In the following Section we report the result obtained from
simulations performed with the random increment distribution given by
Eq.~\equ{eq:gaussian}.  Other sets of simulations, done with a
different peaked distribution [such as an exponential $\rho(\xi)=\exp(-\xi /
a \sigma) /a \sigma$], yield qualitatively similar results, confirming the fact that
the important point is the presence of correlations, and not its
particular form.

\begin{figure}[t]
 \centerline{\epsfig{file=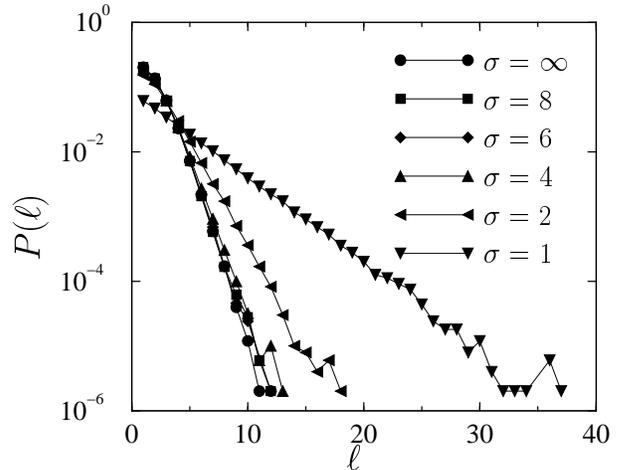,width=8cm}}
 \caption{Chain-length density function $P(\ell)$ for different values of
   the parameter $\sigma$ in $1+1$ dimensional correlated adsorption.}
 \label{fig:size1d}
\end{figure}

\section{Results}

In order to check the effects of correlations in the adsorption
process, we have performed extensive numerical simulations of our
model in $1+1$ and $2+1$ dimensions. Simulations were done in systems
of different size $L$. Statistical averages were performed over at
least $100$ different realizations. In order to avoid cross-effects
between the characteristic length $a \sigma$ and the system size $L$, we
always consider $a \sigma \ll L$. 

\subsection{Adsorption in $1+1$ dimensions}

\begin{figure}[t]
  \centerline{\epsfig{file=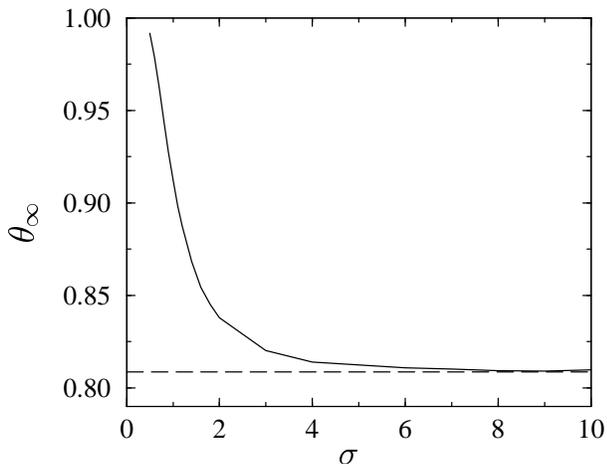,width=8cm}}
  \caption{Jamming limit $\theta_\infty$ as a function of the parameter
    $\sigma$ in $1+1$ dimensional correlated adsorption. The dashed line
    signals the jamming limit for free BM adsorption, $\theta_\infty^{\rm BM}\simeq
    0.80865$.}
  \label{fig:jamming1d}
\end{figure}

We have performed simulations of the correlated adsorption model in
$1+1$ dimensions over a substrate consisting of a line of length
$L=5000a$, with periodic boundary conditions. 

In $1+1$ dimensions, the most relevant feature of the model is the
increased tendency of the adsorbed particles to form connected
structures, identified with chains, whose average length tends to
increase with decreasing $\sigma$. We can use as a measure of this kind of
order the chain-length density function, $P(\ell)$, defined as the
average number of chains of length $\ell a$ per unit length of substrate.
In Fig.~\ref{fig:size1d} we represent the chain-length density
function computed at different values of $\sigma$. The value $\sigma=\infty$
corresponds to simulations of free BM adsorption. For large $\sigma$
we observe a very fast (super-exponential) decay of $P(\ell)$, indicative
of a lack of any characteristic length scale. For small values of $\sigma$,
on the other hand, $P(\ell)$ shows a clear exponential tail, $P(\ell) \sim
\exp(-\ell/ \ell_c)$, with a characteristic length $\ell_c$ depending on $\sigma$. In
particular, for $\sigma=1$ we estimate a value $\ell_c=2.48\pm0.05$.

Another quantity of interest is the jamming limit $\theta_\infty$, which in the
present case can be related to the chain-length density function
through the expression
\begin{equation}
  \label{eq:theta}
  \theta_\infty = \sum_{\ell=1}^\infty  \ell P(\ell).
\end{equation}
In Fig.~\ref{fig:jamming1d} we represent the jamming limit as function
of the parameter $\sigma$. For values of $\sigma$ larger than about $8$ we
recover with good accuracy the limit of free BM adsorption, $\theta_\infty^{\rm
  BM}\simeq 0.80865$ \cite{talbot92}. For smaller values of $\sigma$, on the
other hand, the presence of the correlations induces a higher
compaction on the substrate, with a jamming limit that approaches the
close-packing for a value of $\sigma$ equal to $0.5$ ($\theta_\infty=0.9918\pm0.0005$).

\begin{figure}[t]
  \centerline{\epsfig{file=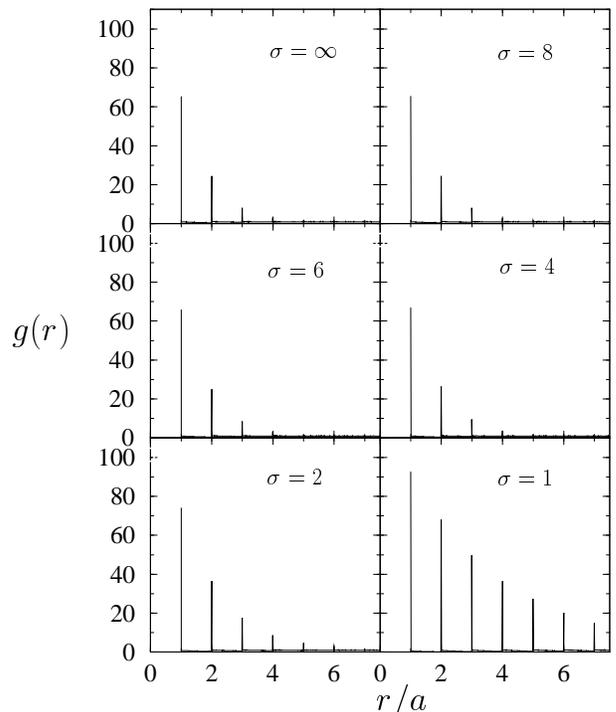,width=8cm}}
  \caption{Particle-particle correlation function $g(r)$ for different
    values of $\sigma$ in $1+1$ dimensional correlated adsorption.}
  \label{fig:correls1d}
\end{figure}

As a final probe of the substrate's structure, we have measured the
particle-particle pair correlation function $g(r)$. In
Fig.~\ref{fig:correls1d} we plot $g(r)$ as a function of the reduced
distance $r/a$ corresponding to six different values of $\sigma$. We
observe that the correlations enhance the maxima of $g(r)$, which
occur at distances $p a$ for integer $p$ (corresponding to
interparticle distances equal to a multiple of the diameter). The
decay of the maxima is close to exponential again. For large values of
$\sigma$ we recover the behavior of the free BM adsorption, with peaks
decaying super-exponentially.

\subsection{Adsorption in $2+1$ dimensions}

\begin{figure}[t]
  \centerline{\epsfig{file=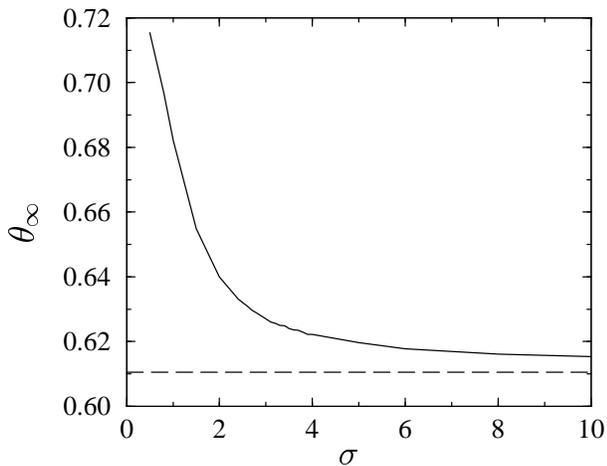,width=8cm}}
  \caption{Jamming limit $\theta_\infty$ as a function of the parameter
    $\sigma$ in $2+1$ dimensional correlated adsorption. The
    dashed line signals the jamming limit for free BM adsorption,
    $\theta_\infty^{\rm BM}\simeq 0.610$.}
  \label{fig:jamming2d}
\end{figure}

Simulations in $2+1$ dimensions were performed on squares of size
$L=125a$, with periodic boundary conditions.  In
Fig.~\ref{fig:jamming2d} we plot the jamming limit $\theta_\infty$ obtained from
simulations for different values of $\sigma$. The trend previously observed
in $1+1$ dimensions is here recovered. For large values of $\sigma$ the
jamming limit tends to the free BM adsorption $\theta_\infty^{\rm BM}\simeq 0.610$
\cite{thompson92}. For decreasing $\sigma$, $\theta_\infty$ increases approaching the
close-packing regime; for $\sigma=0.5$, for example, we estimate
$\theta_\infty=0.715\pm0.001$ \cite{notepack}. It should be noted that the
convergence of $\theta_\infty$ towards its maximum value is slower in this case
than in the adsorption in $1+1$ dimensions (compare
Figs.~\ref{fig:jamming1d} and ~\ref{fig:jamming2d}.).

\begin{figure}[t]
  \centerline{\epsfig{file=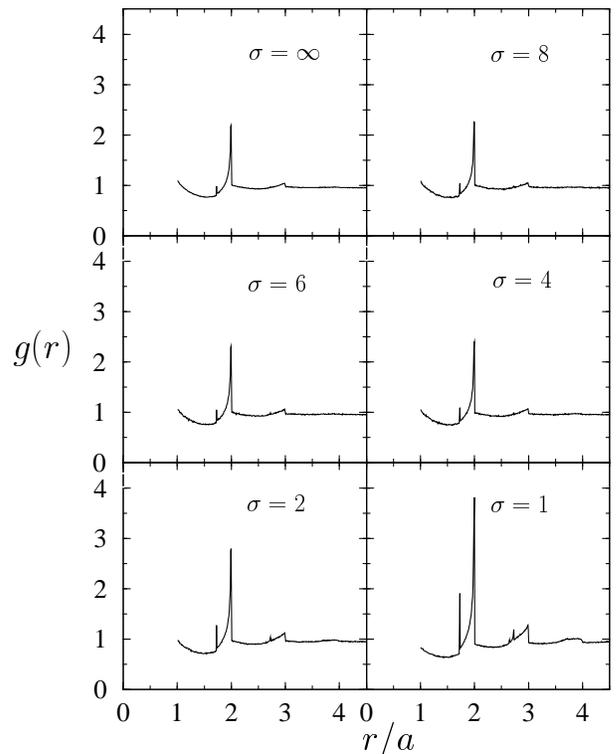,width=8cm}}
  \caption{Particle-particle correlation function $g(r)$ for different
    values of  $\sigma$ in $2+1$ dimensional correlated
    adsorption. The divergent peak at $r=a$ has been suppressed for
    clarity.} 
  \label{fig:correls2d}
\end{figure}

In Fig.~\ref{fig:correls2d} we plot the particle-particle correlation
function $g(r)$ corresponding to the $2+1$ case. For $\sigma=\infty$, we recover
the expected form in free BM adsorption (the divergent peak at $r=a$,
corresponding to the close contact of particles, has been omitted for
the sake of clarity). The secondary peak at $r=2a$ is clearly visible,
as well an intermediate, much smaller peak, located approximately at
$r\simeq 1.73 a$. For decreasing $\sigma$ the height of the secondary peak
increases notably, and the intermediate peak becomes more and more
noticeable.

\begin{figure}[t]
  \centerline{\epsfig{file=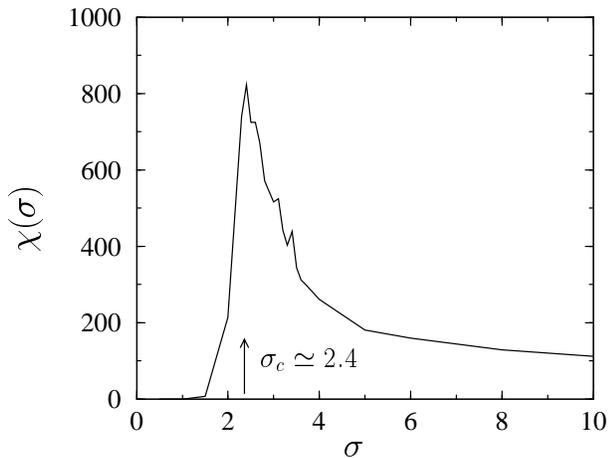,width=8cm}}
  \caption{Susceptibility for the $2+1$ dimensional correlated
    adsorption. The peak, located at $\sigma_c \simeq 2.4$, signals the presence
    of a percolation transition.}
  \label{fig:chi2d}
\end{figure}

\begin{figure}[t]
  \centerline{\epsfig{file=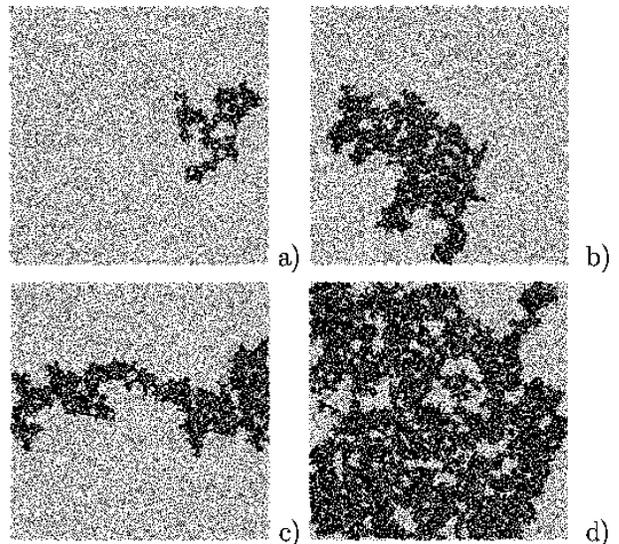, width=8cm}}
  \caption{Largest cluster of connected particles in  $2+1$
    dimensional correlated adsorption. a) $\sigma=\infty$; b) $\sigma=2.8$; c)
    $\sigma=2.4$, d) $\sigma=2.0$.}
  \label{fig:span}
\end{figure}

From the analysis of the correlation function we can conclude that the
correlations in the adsorption mechanism induce the creation of
connected structures on the substrate, which are responsible for the
enhancement of the peaks in $g(r)$. These connected structures cannot
be characterized in terms of chains, since in $2+1$ dimensions
particles adsorb in multiply connected configurations, and therefore
the notion of chain loses its meaning.  In this case, however, the
structure of the adsorbed phase can be better analyzed in terms of its
percolation properties \cite{stauffer94,choi95,vandewalle00}. Since
the particles can roll one above other before adsorption, the
substrate becomes eventually composed by clusters of connected
particles, which can be easily identified.  In free BM adsorption, the
saturated phase has low connectivity and it is below the percolation
threshold: in simulations on any finite system of size $L$, the
largest cluster of connected adsorbed particles is not a spanning
cluster. In order to investigate the percolation properties of the
correlated adsorption model, we have computed the susceptibility $\chi$
of the distribution of clusters of connected particles as a function
of $\sigma$. The susceptibility is defined by \cite{bub90}
\begin{equation}
  \label{eq:suscep}
  \chi= {\sum_s}' n_s s^2,
\end{equation}
where $n_s$ denotes the density of clusters of size $s$ and the prime
in the summation indicates that the biggest (spanning) cluster is
excluded in the sum. At the percolation threshold $\sigma_c$, the
susceptibility is expected to exhibit a peak, with power-law decay
both below and above threshold. In Fig.~\ref{fig:chi2d} we have
plotted the susceptibility $\chi$ computed as a function of the parameter
$\sigma$. The presence of a maximum in $\chi$, located approximately at $\sigma_c \simeq
2.4$, signals the presence of a percolation transition.  Below the
threshold, the clusters of connected particles are numerous and rather
small. Above the threshold ($\sigma<\sigma_c$), however, there exists an
spanning cluster, that crosses the system from boundary to boundary.
We show the presence of such a spanning cluster in
Fig.~\ref{fig:span}. For $\sigma>2.4$, the largest cluster in a typical
simulation of our model is rather small. For $\sigma=2.4$ a largest cluster
develops that first spans the whole system. For even smaller values of
$\sigma$, the largest cluster covers most of the substrate.

\section{Conclusions}

In this paper we have presented a new model of correlated sequential
adsorption of particles on a substrate which takes into account the
existence of certain degree of correlation between the final positions
of the adsorbing particles and the location of the preadsorbed ones.
Its implementation is carried out by modifying the rules of the BM, by
imposing that the final position of an incoming particle is randomly
distributed, with a specified probability density, around the location
of a particle in the substrate. In this way, the model can describe many
different physical situation ranging from the absence of correlations,
corresponding to BM model, to the pure deterministic case in which the
distribution of particles practically reaches the close-packing
configuration.

The main consequence of the existence of correlations is the
appearance of long-range order in the adsorbed phase.  In $1+1$
dimensions, this feature manifests in the fact that the length of the
chains increases when decreasing the variance of the probability
distribution. In $2+1$ dimensions, we have reported the existence of a
percolation transition which can be identified due to the presence of
a peak in the susceptibility. In all cases, the jamming limit
increases significantly when the probability distribution function
becomes sharper, and is close, in the limit of a sharply peaked
distribution, to the close-packing regime.
 
Our model could caricature some real situations, not addressed by the
sequential adsorption models proposed up to now, sharing in common the
fact that the final position of an adsorbed particle may exert an
influence in the location of the next particle. Cases in which the
particles interact, when the kinetics exhibits memory effects or, in
general, when the adsorption process may be in some way controlled by
an external agent and therefore is not totally random, could
accordingly find a description in the model we have proposed.  The
results we have presented may provide new possibilities to the already
existing ones about modelization of adsorption kinetics by sequential
processes.

\acknowledgments
\section*{}

R.P.S. acknowledges support from the grant CICyT PB97-0693. J.M.R.
acknowledges support from the grant DGICyT PB98-1258.


\begin{references}

\bibitem{bartelt91}
M.~C. Bartelt and V. Privman, Int. J. Mod. Phys. B {\bf 5},  2883  (1991).

\bibitem{evans93}
J. Evans, Rev. Mod. Phys. {\bf 65},  1281  (1993).

\bibitem{renyi63}
A. R\'{e}nyi, Sel. Trans. Math. Stat. Prob. {\bf 4},  203  (1963).

\bibitem{feder80}
J. Feder, J. Theor. Biol. {\bf 87},  237  (1980).

\bibitem{schaaf88}
P. Schaaf and H. Reiss, J. Chem. Phys. {\bf 92},  4824  (1988).

\bibitem{senger91}
B. Senger, J.-C. Voegel, P. Schaaf, A. Johner, A. Schmidt, and J. Talbot, Phys.
  Rev. A {\bf 44},  6926  (1991).

\bibitem{ramsden93}
J.~J. Ramsden, Phys. Rev. Lett. {\bf 71},  295  (1993).

\bibitem{meakin87}
P. Meakin and R. Jullien, J. Phys. (Paris) {\bf 48},  1651  (1987).

\bibitem{talbot92}
J. Talbot and S.~M. Ricci, Phys. Rev. Lett. {\bf 68},  958  (1992).

\bibitem{jullien92}
R. Jullien and P. Meakin, J. Phys. A {\bf 25},  L189  (1992).

\bibitem{thompson92}
A.~P. Thompson and E.~D. Glandt, Phys. Rev. A {\bf 46},  4639  (1992).

\bibitem{senger93}
B. Senger, R. Ezzeddine, F.~J. Bafaluy, P. Schaaf, F.~J.~G. Cuisinier, and
  J.-C. Voegel, J. Theor. Biol. {\bf 163},  457  (1993).

\bibitem{schaaf95}
P. Schaaf, P. Wojtaszczyk, B. Senger, J.~C. Voegel, and H. Reiss, Phys. Rev. E
  {\bf 51},  4292  (1995).

\bibitem{pastor98a}
R. Pastor-Satorras and J.~M. Rub\'{\i}, Phys. Rev. Lett. {\bf 80},  5373
  (1998).

\bibitem{adamczyk96}
Z. Adamczyk and P. Warszy\'{n}ski, Adv. Colloid Interface Sci. {\bf 63},  41
  (1996).

\bibitem{notecorrelations}
The BM in its original formulations, was referred to as ``correlated sequential
  adsorption'' (CSA) deposition, due to the very short-range correlations among
  particles induced by the sliding mechanism incorporated in the model
  \cite{thompson92}. In the present paper we are concerned with correlations of
  a more general nature, induced by external agents or interactions.

\bibitem{viot93}
P. Viot, G. Tarjus, and J. Talbot, Phys. Rev. E {\bf 48},  480  (1993).

\bibitem{choi95}
H.~S. Choi, J. Talbot, G. Tarjus, and P. Viot, Phys. Rev. E {\bf 51},  1353
  (1995).

\bibitem{notepack}
The maximum value of $\theta_\infty$ in $2+1$ dimensions will be smaller than
  the one corresponding to a close-packed substrate, due to the inherent
  randomnes of our model.

\bibitem{stauffer94}
D. Stauffer and A. Aharony, {\em Introduction to Percolation Theory}, 2on ed.
  (Taylor \& Francis, London, 1994).

\bibitem{vandewalle00}
N. Vandewalle, S. Galam, and M. Kramer, Eur. Phys. J. B {\bf 14},  407  (2000).

\bibitem{bub90}
S.~B. Lee, Phys. Rev. B {\bf 42},  4877  (1990).

\end{references}
\end{document}